\documentclass[pra,aps,twocolumn]{revtex4}
\usepackage{epsf,amsmath,color,graphicx}
\newcommand {\be}{\begin{equation}}
\newcommand {\ee}{\end{equation}}
\newcommand {\bea}{\begin{eqnarray}}
\newcommand {\eea}{\end{eqnarray}}

\begin{document}

\title{Atomtronics: ultracold atom analogs of electronic devices}
\author{B. T. Seaman, M. Kr\"amer, D. Z. Anderson and M. J. Holland}
\affiliation{JILA, National Institute of Standards and Technology and
Department of Physics, \\University of Colorado, Boulder CO
80309-0440, USA} \date{\today}

\begin{abstract}
{\it Atomtronics} focuses on atom analogs of electronic materials,
devices and circuits.  A strongly interacting ultracold Bose gas in a
lattice potential is analogous to electrons in solid-state crystalline
media.  As a consequence of the band structure, cold atoms in a
lattice can exhibit insulator or conductor properties.  P-type and
N-type material analogs can be created by introducing impurity sites
into the lattice.  Current through an atomtronic wire is generated by
connecting the wire to an atomtronic battery which maintains the two
contacts at different chemical potentials.  The design of an
atomtronic diode with a strongly asymmetric current-voltage curve
exploits the existence of superfluid and insulating regimes in the
phase diagram.  The atomtronic analog of a bipolar junction transistor
exhibits large negative gain.  Our results provide the building blocks
for more advanced atomtronic devices and circuits such as amplifiers,
oscillators and fundamental logic gates.


\end{abstract}

\maketitle

\section{Introduction}
\label{sec:introduction}

A collection of ultracold atoms subject to a spatially periodic
potential can exhibit behavior analogous to electrons in a crystal
lattice.  This fact has been established in an impressive series of
experiments with Bose-Einstein condensates and Fermi gases in optical
lattices, that is, periodic potentials produced by interfering laser
beams
\cite{Anderson1998,Cataliotti2001,Greiner2001b,Morsch2001,Denschlag2002,Modugno2003,Roati2004,Pezze2004,Fort2005,Kohl2005}.
The analogy between ultracold atoms in lattice potentials and
electrons in crystals is manifestly a rich one.  It extends to
strongly interacting ultracold Bose gases which exhibit both
superfluid and insulating behavior and feature a gapped energy
spectrum. The tunability of interactions in optical lattices has led
to the spectacular demonstration of these properties
\cite{Greiner2002,Stoferle2003}. In this work we introduce analogs of
electronic materials, including metals, insulators, and
semiconductors, in the context of ultracold strongly interacting
bosons.  We use lattice defects to achieve behavior similar to doped
P-type and N-type semiconductors. The interest is to adjoin P-type and
N-type lattices to create diodes and then NPN or PNP structures to
achieve behavior similar to that of bipolar junction transistors.  We
show that such heterogeneous structures can indeed be made to mimic
their electronic counterparts. Ruschhaupt and Muga
~\cite{Ruschhaupt2004} have described an atom device with diode-like
behavior, and Micheli et.~al.~\cite{Micheli2004} have proposed a
single-atom transistor that serves as a switch (see also
\cite{Ruschhaupt2005,Ruschhaupt2006a,Ruschhaupt2006b,Daley2005,Micheli2006}).
Both of these devices depend on control and coherence at the single
atom level.  Moreover, Stickney et.~al.~\cite{Stickney2006} have
recently demonstrated that a Bose-Einstein condensate in a triple well
potential can exhibit behavior similar to that of a field effect
transistor.  Our intent is to establish ultracold atom analogs of
electronic materials and semiconductor devices that can be used to
leverage the vast body of electronic knowledge and heuristic methods.
From semiconductor materials, the analogy expands into what can be
referred to as {\it atomtronics}. With diodes and transistors in hand,
it is straightforward to conceive of atom amplifiers, oscillators,
flip-flops, logic gates, and a host of other atomtronic circuit
analogs to electronic circuits.  Such a set of devices can serve as a
toolbox for implementing and managing integrated circuits containing
atom optical elements \cite{Rolston2002,Bongs2004,Search2005} or
quantum computation components
\cite{Brennen1999,Briegel2000,Steane1998} and might be of particular
interest in the context of rapidly advancing atom chip technologies
\cite{Folman2002,Reichel2002}.

Atoms in periodic potential structures and electrons in solid state
crystals have much in common. In both systems, particle motion occurs
by tunneling through the potential barrier separating two lattice
sites.  A particle can delocalize over the entire lattice and sustain
currents.  Therefore, in atomtronic devices the dual of electric
current is atomic flux.  In both systems, currents are created when
there is a potential gradient which causes the particles to move from
a region of high potential to a region of low potential. In
electronics, potentials arise from electrical fields. In atomtronics,
potential gradients are understood in terms of chemical
potentials. The characteristics of both electronic and atomtronic
devices can be examined by using a battery to apply a potential
difference across the system and observing the response in the
current.

Different types of electronic conductors exist because electrons in a
crystal structure occupy states of an energy spectrum that features a
band structure. The materials can carry a current only if the highest
occupied energy band is only partially filled with electrons. Under
this condition, the system is a conductor. On the other hand, the
system is an insulator if all occupied bands are full.  The highest
fully occupied band is called the valence band, while the lowest empty
or partially occupied band is referred to as the conduction band.

\begin{figure}
\begin{center}
\includegraphics[width=7.8cm]{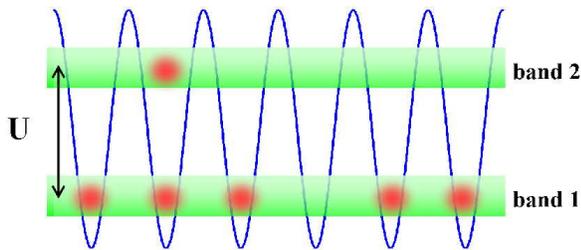}
\caption{Schematic of the atomtronic energy band structure of strongly
interacting bosons in a lattice. The first band is made up of all
states with a filling of less than one atom per site while the second
band contains all states with filling between one and two atoms per
site. The two bands are separated by the onsite interaction $U$.}
\label{fig:undopedbands}
\end{center}
\end{figure}

Ultracold strongly interacting bosonic atoms in periodic structures
behave similarly to their electronic counterparts.  Strong repulsive
interactions prevent atoms from occupying the same lattice site,
mimicking the fermionic behavior of electrons. Hence, a current can
flow easily as long as there are empty sites available.  However, once
the filling reaches one atom per site, the system becomes an
insulator. A large energy gap given by the repulsive onsite
interaction must be overcome in order to add another particle to this
configuration. A particle added in excess of a filling of one atom
per site can again carry a current since it can move around freely
above the filled valence band of one atom per site. The system remains
a conductor until it arrives at a filling of two atoms per site and it
becomes an insulator again.  Hence, as in electronics, there is a 
band structure and it is the filling of the bands which determines
whether a material is a conductor or an insulator. The energy band
structure of atoms in a lattice is depicted schematically in
Fig.~\ref{fig:undopedbands}. The lowest band is made up of states
having between zero and one atom per site. The next band contains all
states with one to two atoms per site. This second band is separated
from the first by a large energy gap on the order of the onsite
repulsion. Higher bands are formed analogously. The highest occupied
band of a conductor is only partially filled while insulators are
characterized by full bands.

There are two major differences between usual electronic materials and
the atomtronic materials considered here. First, the energy gap is due
to the Pauli exclusion principle in the case of electrons while in the
bosonic case the gap is due to the repulsive interaction between
atoms. Second, the atomtronic conductor features superfluid rather
than normal flow and in that sense resembles an electronic
superconductor.

The purpose of this paper is to study the behavior of atomtronic
materials in simple circuits and to show how to use these materials to
build more complex circuit devices, namely diodes and bipolar junction
transistors. The paper is structured as follows.  Section
\ref{sec:formalism} introduces the Bose-Hubbard formalism that will be
used to describe the atomtronic systems. In particular, it discusses
the zero temperature phase diagram, the properties of an atomtronic
battery and the method we use to calculate current response to a
voltage.  Doped atomtronic materials and the current-voltage behavior
of atomtronic wires will be discussed in Sec.~\ref{sec:materials}.  A
diode obtained by combining a P-type and N-type atomtronic material
with a voltage bias applied by a battery is examined in
Sec.~\ref{sec:diode}.  Section~\ref{sec:transistor} presents the atom
analog of a semiconductor bipolar junction transistor. Finally,
Sec.~\ref{sec:remarks} contains remarks on possible applications, on
the differences between our atomtronic devices and their electronic
counterparts and on future perspectives.

\section{Bose-Hubbard Formalism}
\label{sec:formalism}

The different kinds of atomtronic materials are best understood by
examining the different zero temperature quantum phases of the system.
We model the system of bosons in a 1D chain of lattice sites with the
Bose-Hubbard Hamiltonian \be \hat{H}=\frac{U}{2}\sum_i
\hat{n}_i(\hat{n}_i-1)-J\sum_{\langle ij\rangle}\hat{a}_i^\dag
\hat{a}_j+\sum_i (\epsilon_i-\mu) \hat{n}_i\, ,
\label{eq:hamiltonian}\ee where $\hat{a}_i$ is the annihilation
operator for a particle at site $i$,
$\hat{n}_i\equiv\hat{a}_i^\dag\hat{a}_i$ is the number operator at
site $i$, $U$ is the onsite repulsive interaction strength, $J$ is the
hopping matrix element between nearest neighbors, $\langle ij \rangle$
labels nearest neighbors, $\epsilon_i$ is the external potential at
site $i$ and $\mu$ is the chemical potential of the system. The
Bose-Hubbard Hamiltonian is obtained by retaining only the
contributions of the lowest single particle Bloch band to the Hilbert
space and by making a tight binding approximation (for a review see
\cite{Zwerger2003}). It yields an accurate description of an ultracold
dilute Bose gas in a periodic potential. The zero temperature phase
diagram of this Hamiltonian was first studied in \cite{Fisher1989}.

For very large onsite repulsion, $U\gg J$, the system enters the
regime of {\it fermionization} where bosons are impenetrable and only
two Fock states, $|n_i\rangle$ and $|n_i+1\rangle$, are needed at each
site to accurately describe the system (two-state approximation).
Note that this is equivalent to mapping bosonic operators onto
fermionic ones via the Jordan-Wigner transformation
\cite{Jordan1928,Sachdev1999}.  The data presented in this paper is
obtained by considering the system described by the Hamiltonian
Eq. (\ref{eq:hamiltonian}) in the fermionized regime. We have verified
that the error resulting from the two-state approximation becomes
negligible for $U/J\ge 100$ (see also \cite{Bhat2006}).

\subsection{Phase Diagram}
\label{subsec:phasediagram}

\begin{figure}
\begin{center}
\includegraphics[width=7.8cm]{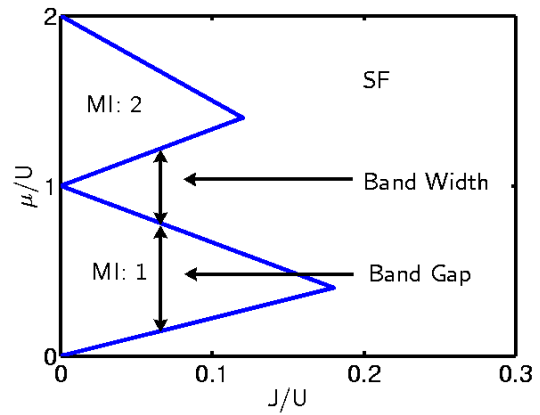}
\caption{Zero temperature phase diagram of a Bose gas obtained for a
one-dimensional lattice using the two-state approximation. At large
values of $J/U$ the gas is superfluid (SF). Below a critical value of
$J/U$ the system enters a Mott-insulator phase (MI) for integer
filling, while it remains superfluid for non-integer filling. The $MI:
1$ Mott-insulator region has one atom per site and the $MI: 2$ region
has two atoms per site. The Mott lobe boundaries are not rounded due
to the use of the two-state approximation \protect{\cite{note}}.}
\label{fig:basicphasediagram}
\end{center}
\end{figure}

The phase diagram of the Bose-Hubbard Hamiltonian contains the
complete information about the band structure of the system for a
given $J/U$, see Fig.~\ref{fig:basicphasediagram} in which the phase
diagram was created using the two-state approximation. At $T=0$, the
Bose-Hubbard model has two distinct phases, a Mott-insulating and a
superfluid phase.  The Mott-insulating phase is entered below a
critical value of $J/U$ for an integer number of particles per site.
In this phase strong interactions completely block particle motion
rendering the gas incompressible, that is $\partial n/\partial \mu
=0$, where $n$ is the average filling of a site. The superfluid phase
is obtained for non-integer filling.  Figure
\ref{fig:basicphasediagram} presents the boundary between the
conducting and insulating phases as a function of $J/U$ and chemical
potential $\mu/U$.  The two lobes limit the Mott-insulator zones with
one atom per site (lower lobe, MI:1) and two atoms per site (upper
lobe, MI:2).  Insulator phases with higher integer filling are
obtained at larger values of $\mu/U$. The remainder of the phase
diagram is in the conducting superfluid phase, labeled $SF$.  No
insulating phase exists for values of $J/U$ larger than $ ~\sim
1$. Note that the triangular, non-rounded, shape of the Mott lobes in
Fig.~\ref{fig:basicphasediagram} is due to the two-state approximation
becoming increasingly inaccurate as $J/U$ is increased \cite{note}.

Each value of $\mu/U$ and $J/U$ maps onto a particular lattice
filling. In the fermionization regime ($J/U\ll 1$), an analytic
expression for the relation between these parameters in the limit of a
large number of lattices sites can be derived. It reads
\cite{Peden2006} \be \mu = U(m-1)+(-1)^m 2 m J \cos(\pi n)\,,
\label{eqn:mu}\ee
where $m=1,2,\dots$ is the band index. Plotting the phase diagram as a
function of $\mu/U$ rather than the number of particles is useful
because then one can directly read out the band widths and the band
gaps for a given $J/U$. As illustrated in
Fig.~(\ref{fig:basicphasediagram}), the width of the second band is
given by the height of the second superfluid phase region at fixed
$J/U$ while the size of the band gap between first and second band is
given by the height of the lowest insulating zone at that $J/U$.
Hence, the details of the atomic band structure, such as the exact
size of the band gap and height of a band, depend on the ratio
$J/U$. In order to have access to both insulating and conducting
phases, the ratio $J/U$ must be small.  If this ratio is too big there
is no well defined band gap and therefore no transition to an
insulating phase for integer filling.  The basic ideas presented in
this paper rely on this condition being satisfied. They do not require
the stronger condition $J\ll U$ for fermionization. The latter
condition is merely assumed to facilitate calculations.

\subsection{Atomtronic Battery}
\label{subsec:atomicbattery}

Just as for electronic circuits, the atomtronic version of a battery
is crucial but it is also subtle.  In the following, we identify the
basic properties and actions of the atomtronic analog of battery.

Energy for electronic circuits is supplied by sources of electric
potential. Furthermore, electric potentials are used to set the bias
points of circuit elements to achieve their desired behavior. For
simplicity, we will use the term battery to refer to a device that
provides a fixed potential difference and can supply an electric
current or atom flux.

The voltage of a battery is fixed by the potential difference between
two terminals. A specific potential difference between two points
within a device or circuit is achieved by connecting those two points
with the two battery terminals, or poles. In the electronic case the
value of electric potential at any one point in a circuit is
arbitrary, as one is interested only in electric potential
differences, in other words the voltage between two points.  In the
atomtronic case, the function of the battery is to hold the two
contacts at two different values of the chemical potential, say
$\mu_L$ on the left and $\mu_R$ on the right.  The applied voltage is
then defined by \be V\equiv\mu_L-\mu_R\,.\ee The current flows from
higher to lower chemical potential, i.e.~from higher to lower
voltage. Chemical potential difference in atomtronic system is
analogous to electric potential difference in electronic systems.

To understand the physics underlying this concept, note that bringing
the system in contact with a battery pole of chemical potential
$\mu_L>\mu$ leads to the injection of $\Delta n$ particles, where
$\mu$ is the chemical potential of the isolated system. The magnitude
of $\Delta n$ is given by the difference in filling of states with
chemical potential $\mu$ and $\mu_L$. This information is contained in
the phase diagram. The particle transfer increases with increasing
$\mu_L$ within a superfluid region and becomes constant as $\mu_L$ is
moved into a Mott insulating zone.  In the fermionization regime, the
magnitude of $\Delta n$ is fixed by Eq.~(\ref{eqn:mu}).  The analogous
reasoning applies to the removal of particles at the battery pole with
$\mu_R<\mu$.

Feeding atoms into a circuit element through a contact at one end and
removing them through a contact at the other end generates a current.
Particles move from a region of excess to a region of deficit. This
current reaches a steady state if the carrier excess at one end is
replenished through one contact with the battery at the same rate at
which the deficit is maintained at the other end through the contact
with the other pole of the battery. 

\begin{figure}
\begin{center}
\includegraphics[width=7.8cm]{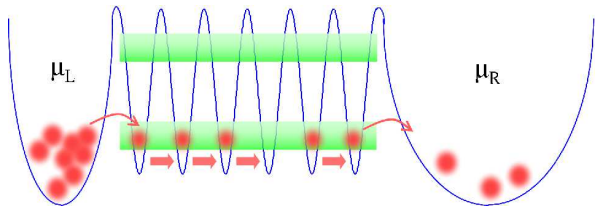}
\caption{Schematic of atoms in a lattice connected to an atomtronic
battery. A voltage is applied by connecting the system to two
reservoirs, one of higher (left) and one of lower chemical potential
(right). An excess (deficit) of atoms is generated in the left (right)
part of the system respectively giving rise to a current from left to
right.
\label{fig:battery}}
\end{center}
\end{figure}

Experimentally, a battery can be created by establishing two separate
large systems which act as reservoirs each with its own constant
chemical potential.  These may also be lattice configurations or other
experimentally plausible systems such as large harmonic traps
containing a large number of atoms.  Each of these reservoirs can be
connected to one end of the atomtronic system and current is then
possible from the higher chemical potential system to the lower one.
This configuration is displayed in Fig.~\ref{fig:battery}. From a
practical point of view, the chemical potentials of the battery poles
can be maintained by transferring atoms, possibly classically, between
the two poles.

\subsection{Computation of Current Response}
\label{subsec:batterymodel}

To power a circuit, we connect the system to an atomtronic battery by
bringing two reservoirs of different chemical potential $\mu_L$ and
$\mu_R$ into contact with the two ends of the lattice. The values of
$\mu_L$ and $\mu_R$ determine the particle transfer $\Delta n$ which
is injected through one contact and removed through the other. The
transfer $\Delta n$ follows from the relations $\mu_L(n+\Delta n)$ and
$\mu_R(n-\Delta n)$ with $n$ the filling of the system at zero
voltage. Note that, the chemical potentials $\mu_L$ and $\mu_R$ must
be chosen appropriately to ensure that the average filling is kept at
$n$.

The net current obtained by applying a voltage is given by the average
number of particles passing through the system per unit time after a
steady state situation has been reached.  Yet, the current not only
depends on the particle transfer $\Delta n$ as determined by the
values of $\mu$, $\mu_L$ and $\mu_R$, but also on the strength of the
coupling between battery and system.  The latter is set by the
properties of the battery contact and determines the rate at which
transfers of magnitude $\Delta n$ take place. The maximum current is
obtained when subsequent transfers are separated by just the time it
takes for one transfer of $\Delta n$ to free up the site to which the
battery is connected. This is the regime we focus on because it yields
the largest currents and reveals the limits on the currents which are
due to the properties of the conductor material rather than due to the
properties of the battery contact.

The steady state reached after a time of transient behavior takes the
form of a running density wave whose amplitude is determined by the
particle transfer $\Delta n$ from the battery, hence by the voltage,
and whose wavelength depends on the strength of the coupling between
battery and wire.  The maximum current is present when the
wavelength is given by twice the lattice period. In that case, the
maxima (minima) of the wave take the value $n+\Delta n$ ($n-\Delta
n$).  The steady state current is then essentially given by $I=4\Delta
n\nu$, with $\nu$ the frequency of the wave, since the time it takes
to transport $2\Delta n$ particles through a single link is given by
$\Delta t=1/2\nu$. To give an example, in the simplest case of a
single link in the absence of additional external potentials, it is
given by $\Delta t = 1/2\nu$ where $\nu$ is the Rabi frequency
$\nu=J/\hbar\pi$. In the limit of small voltages, the wave created by
the contact with the battery can be described by the elementary
excitations of the gas.

We estimate the steady state currents by considering systems which
only have a few links.  The idea underlying our calculation of the
current response is that it mainly depends on the particle transfer
$\Delta n$ associated with a certain voltage and only negligibly on
the number of lattice sites. Reducing the calculation down to only one
or two links is possible since we are interested in the regime $J/U\ll
1$ where beyond next-neighbor correlations are of minor importance.

In the case of atomtronic wires and diodes we consider a single link
because in these cases the steady state current is controlled by one
type of link. While in a wire there is only one type of link, the
current in a diode is controlled by transport through the link at the
junction between P-type and N-type material. A two-link system needs
to be considered to describe a transistor since in this case the
current is controlled by both the links at the interfaces between base
and collector and between base and emitter. Current in wires, in a
diode and a transistor will be discussed in detail below.

In practice, we compute the dynamical evolution of the one- or
two-link system we are interested in. The system is prepared in a
non-stationary state corresponding to the configuration generated by
the battery contact. The actual transfer of particles between system
and battery is not part of the calculation. The role of the battery is
taken into account in the choice of the initial state which involves
an excess $\Delta n$ on one side and a deficit $\Delta n$ on the
other. We determine the steady state current $2\Delta n/\Delta t$
where $\Delta t$ is the time it takes for $2\Delta n$ particles to
move from the initially higher populated site to its neighbor.

The lattices necessary to actually build atomtronic diodes and
transistors must be large since their operation relies on the
existence of superfluid and insulating phases, i.e. on the
applicability of the phase diagram
Fig.~\ref{fig:basicphasediagram}. The importance of the phase diagram
lies in determining the relation between chemical potential $\mu$ and
filling $n$.  As explained above, it is this relation which fixes the
particle transfer $\Delta n$ between battery terminal and atomtronic
system at a certain voltage. Above all, it determines the range of
voltages in which a change of voltage does not bring along a change in
current because the battery chemical potentials lie in insulating
zones where $\partial n/\partial\mu=0$.  Once we have obtained the
data for the steady state current as a function of $\Delta n$, we use
the large-system relation Eq.~(\ref{eqn:mu}) for $\mu(n)$ to represent
the current as a function of chemical potential difference,
i.e. voltage.

\section{Atomtronic Conductors}
\label{sec:materials}

The conductivity properties of the material change with the number of
atoms in the lattice, i.e. with the lattice filling. The filling is
increased as the chemical potential is raised. This moves the point in
the phase diagram and yields either a superfluid or an insulating
state (see Fig.~\ref{fig:basicphasediagram}). Changing the chemical
potential is not the only way to control the filling and thus the
properties of the material.  In this section, we first discuss the
possibility of varying the conductivity of a material by lattice
doping. Secondly, we focus on the current-voltage characteristics of
the different types of atomtronic materials.

\subsection{Doped Materials}
\label{subsec:doping}

\begin{figure}
\begin{center}
\includegraphics[width=7.8cm]{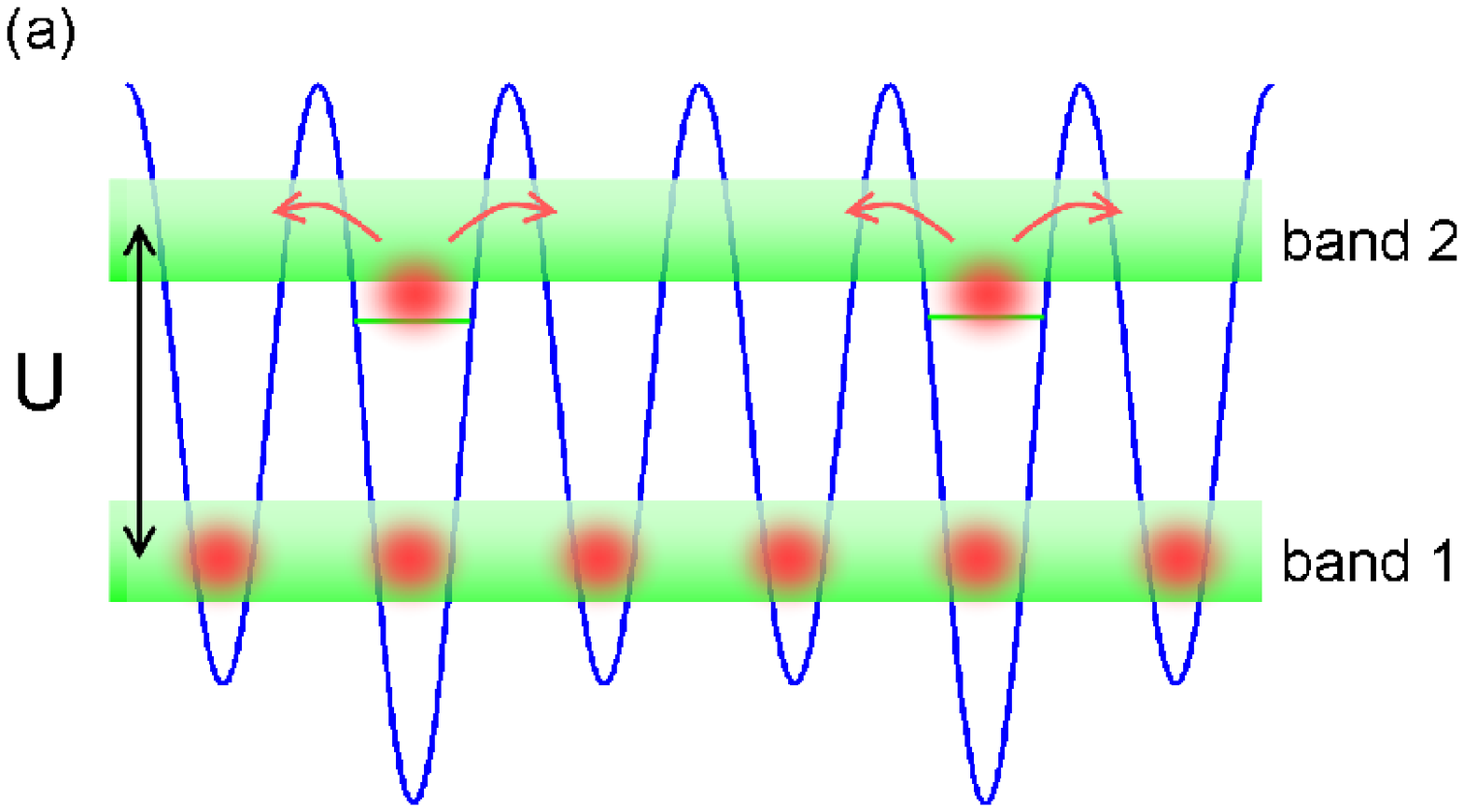}\\~\\
\includegraphics[width=7.8cm]{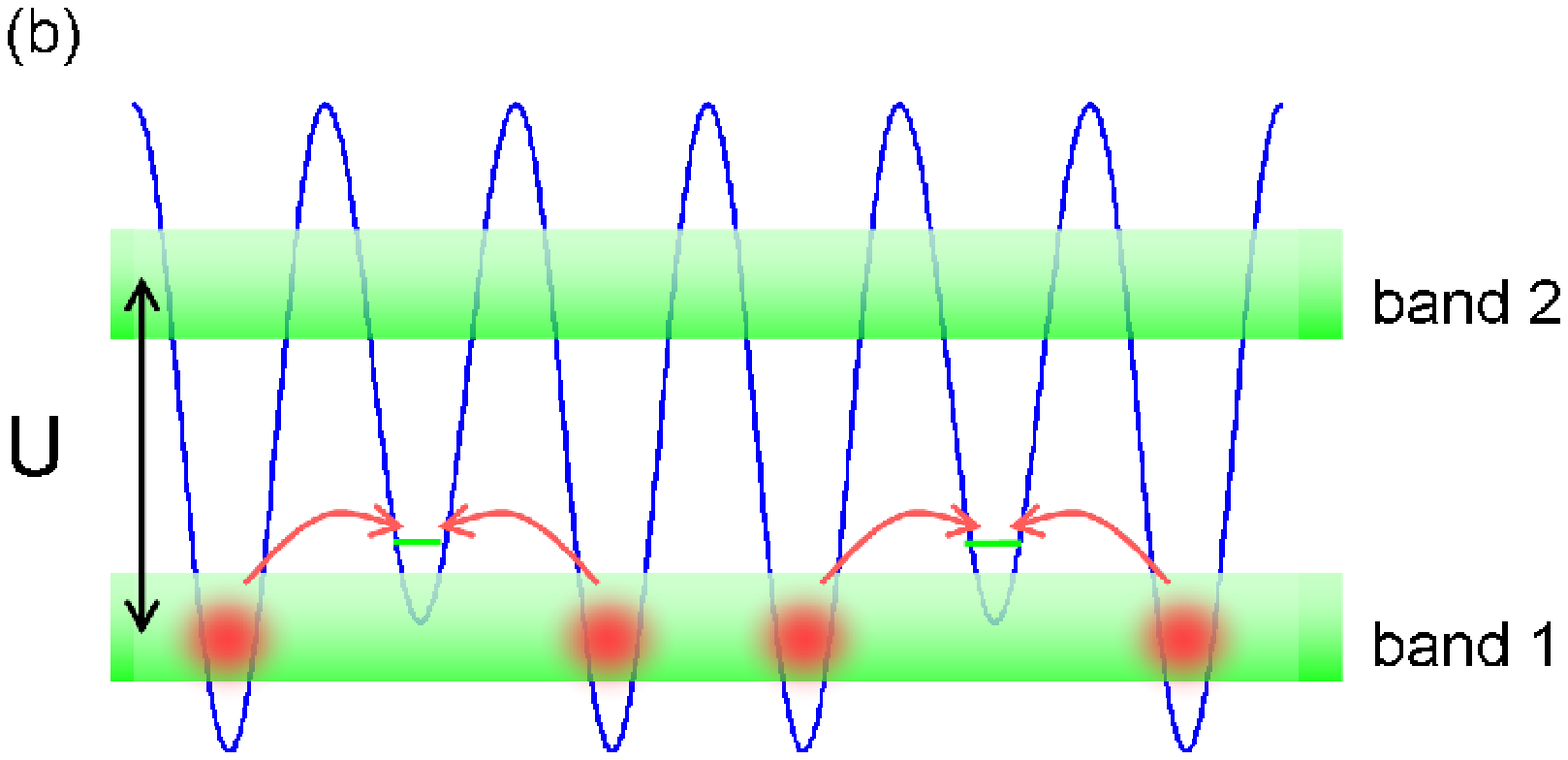}
\caption{ (a) Schematics of an N-doped lattice. The donor sites
feature a level right underneath the first empty band. An atom which
occupies this level can easily be excited and move throughout the
lattice.  \\ (b) Schematics of a P-doped lattice. Acceptor sites have
a level right above the highest full band. Atoms can easily be excited
into this level and allow for a hole to move throughout the lattice.
\label{fig:doping}}
\end{center}
\end{figure}

New materials with interesting properties can be designed by modifying
the lattice potential in which the atoms are confined, and hence the
band structure. This can be done in various ways. In the case of an
optical lattice, the periodicity of the lattice can be modified by
superimposing a periodic potential variation of different wavelength
\cite{Peil2003}. Disorder can be introduced by randomly modifying
individual sites \cite{Horak1998,Gimperlein2005,Schulte2005}. A
further handle on the properties of the material is the symmetry of a
2D or 3D lattice \cite{Grynberg2001,Santos2004,Sanchezpalencia2005}.
Finally, it is conceivable to introduce a second atomic species,
bosonic or fermionic, and to modify the properties of the material via
inter-species interaction.

The atom analog of the doping of a semiconductor is particularly
interesting. The aim of doping is to create energy levels in the
energy gap between two bands. P-type doping is associated with energy
levels located close to the upper edge of the highest full band while
N-type doping gives rise to levels close to the lower edge of the
first empty band. Both are accomplished by modifying the potential at
individual lattice sites.  N-type doping is achieved by replacing some
lattice sites with donor sites. These correspond to potential wells
which are slightly deeper than those of the unmodified
lattice. Analogously, P-type doping requires introducing acceptor
sites of slightly shallower potential. The two potential
configurations are shown in Fig.~\ref{fig:doping}. 

\begin{figure}
\begin{center}
\includegraphics[width=7.8cm]{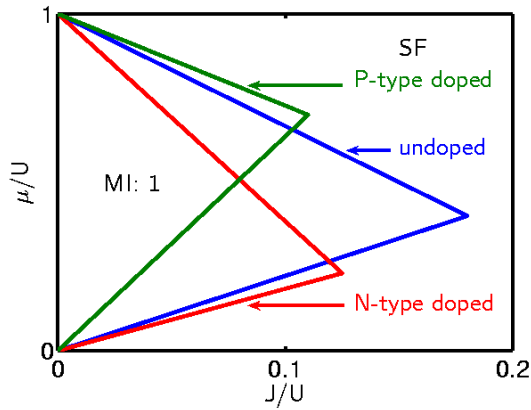} 
\caption{Zero temperature phase diagram of a Bose gas in an undoped
(blue), an N-doped (red) and a P-doped lattice (green). Displayed is
the boundary of the lowest Mott insulator lobe with a filling of one
atom per site as calculated using the two-state approximation. The
results for the doped lattices were obtained by adding/subtracting an
energy of $5J$ to/from every third site in a lattice with six sites.
N-type doping turns an insulating state into a superfluid state with
more than one atom per site. Similarly, P-type doping turns an
insulating state into a superfluid state with less than one atom per
site.}
\label{fig:dopingphasediagram}
\end{center}
\end{figure}

The advantage of doping is that it allows turning an insulator into a
conductor without having to excite particles across the band gap into
the empty conduction band. In atomtronics, this means that doping
shifts the insulator zone in the phase diagram. Figure
\ref{fig:dopingphasediagram} compares the phase diagram of the undoped
lattice with that of a P-doped and a N-doped lattice.  N-type doping
shifts the insulating zone downwards such that states that were
previously in the insulating zone come to lie right above the
insulating zone where the lattice has a full valence band and a few
free carriers in the conduction band. Similarly, P-type doping shifts
the insulating zone upwards such that states previously in the
insulating region come to lie right below the insulating zone where
the lattice has an almost full valence band with a few free hole
carriers.

\subsection{Material Current Properties}
\label{subsec:materialcurrentproperties}

This section discusses the current response of atomtronic wires to an
applied voltage.  The magnitude of the current depends on the
properties of the material and on the nature of the contact with the
battery. The material can be in the Mott insulating phase or in the
superfluid phase. In the Mott phase, a small voltage will not yield a
current. Only at a large voltage, of the order of the gap $U$, will
the battery be able to generate a current by feeding particles into
the next unoccupied band. If the material is in the superfluid phase
even a small voltage will suffice to generate a current. Due to the
superfluid nature of the atomic carriers, this current is not slowed
down by friction. Hence, the ratio of voltage to current does not have
the physical meaning usually associated with resistance, but reflects
the limits on the current at a given voltage due to factors other than
decelerating friction. A primary limit is set by the hopping parameter
$J$ which quantifies how quickly atoms can move from one site to the
next. Its value depends on the shape and the depth of the lattice. A
further limit on the current is due to the interaction between the
atoms. Their repulsion forces them to move in a highly correlated
fashion. For this reason, the current does not grow linearly with the
number of carriers at fixed voltage. Instead, the current per particle
drops as more carriers are added and eventually goes to zero when the
filling is one atom per site and the system becomes an insulator.

Apart from material dependent factors, it can be the nature of the
contact with the battery which fixes the maximum achievable
current. If it is more difficult for atoms to pass through the contact
than to hop from one lattice site to the next then the current is not
limited by $J$ but by the rate at which the battery feeds in and
removes particles. This situation is encountered when operating a
battery in a regime of very weak coupling. The same properties are
present in electronic systems where there can be different types of
contacts, such as rectifying and ohmic contacts~\cite{Ankrum1971}.
Since we focus on the regime of maximum currents, currents are limited
by the hopping parameter $J$ and are not influenced by the properties
of the battery contact.

\begin{figure} \begin{center}
\includegraphics[width=7.8cm]{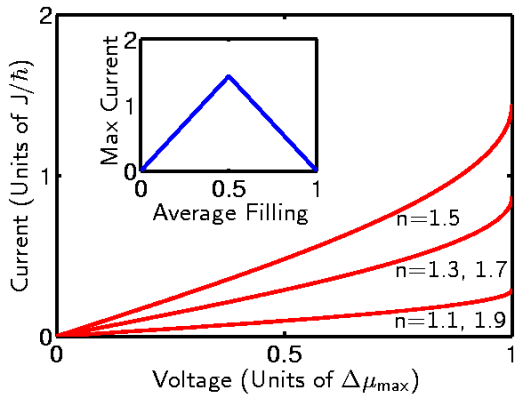}
\caption{Current as a function of chemical potential difference
(voltage) for different materials (different fillings $n$) in the
fermionization regime where bosons are impenetrable.  A wire with a
certain number of atoms carries the same current as a wire with that
number of holes (particle-hole symmetry).  The chemical potential
difference is given in units of $\Delta\mu_{\rm max}$. This quantity
denotes the chemical potential differences that yield the maximum
currents and corresponds to the chemical potential difference at which
the system enters an insulating regime at one of the two battery
contacts for a given $n$.  Inset: Current as a function of filling at
$\Delta\mu_{\rm max}$. Maximum currents are attained at half
filling. Note the symmetry in the current due to the particle-hole
symmetry.
\label{fig:ivcurveflat}}
\end{center} \end{figure}

As examples of the current-voltage characteristics of atomtronic
wires, Fig.~\ref{fig:ivcurveflat} presents the current as a function
of voltage for lattices, or wires, of average filling $n=1.1, 1.3,
1.5, 1.7, 1.9$. The curves for fillings with an equal number of free
particles and holes coincide demonstrating the equivalence of hole and
particle motion for $J/U\ll 1$. Since the gas is superfluid, the
attainable currents are limited by the hopping parameter $J$ and not
by a dissipative mechanism that in electronic systems gives rise to a
resistance.  The maximum current attainable for a half-filled second
band is $\sim 1.4\,J/\hbar$. To give an example, note that in a cubic
optical lattice with a depth of $10\,E_R$ the hopping parameter $J\sim
0.02 E_R$, where $E_R=\hbar^2\pi^2/2md^2$ is the recoil energy which
is fixed by the lattice period $d$. For $87Rb$ and a lattice period of
$d=400 nm$, the recoil energy is $E_R=(2\pi\hbar)3.55 kHz$ yielding
currents on the scale $J/\hbar\sim 2\pi 71 Hz$ in a one-dimensional
lattice.

The voltage in Fig.~\ref{fig:ivcurveflat} is given in units of
$\Delta\mu_{\rm max}$.  This quantity denotes the chemical potential
differences that yields the maximum currents.  It corresponds to the
chemical potential difference at which the system, for a given
filling, enters an insulating regime at one of the two battery
contacts. 

To better understand the current carrying characteristics of the
different materials, the inset in Fig.~\ref{fig:ivcurveflat} presents
the currents at different average fillings, corresponding to different
materials, at $\Delta\mu=\Delta\mu_{\rm max}$.  The plot is symmetric
around its maximum at half filling reflecting particle-hole symmetry.
Note that the average current per free particle, i.e. holes at
$n>1/2$, is constant. These current characteristics are very similar
to those of semiconductor materials in that metals are good conductors
while doped semiconductors do not conduct as well.

The data displayed in Fig.~\ref{fig:ivcurveflat} is obtained from the
dynamical evolution of a single link within the two-state
approximation. The initial state is prepared with $n_R$ and $n_L$
particles on the right and left respectively. This is the state
generated by applying a voltage $V=\mu_L-\mu_R$ with $\mu_L=\mu(n_L)$
and $\mu_R=\mu(n_R)$ where in general the relation between chemical
potential and filling is obtained from the phase diagram and is given
by Eq.~(\ref{eqn:mu}) in the fermionized regime. The evolution of this
initial state features the time-dependent current \be
i(t)=\frac{2J}{\hbar}(n_R-n_L)\;\sin(4Jt/\hbar)\,
\label{eq:current1}\ee 
from right to left. To obtain the steady state current we maximize the
time average of Eq.~\ref{eq:current1} over a time interval $\Delta
t$. This yields \be I=1.44 \frac{J}{\hbar}(n_R-n_L)\ee with $\Delta
t=0.83 \hbar/J$. This value is close to the time it takes to invert
the initial population imbalance, i.e. to complete half a cycle of
frequency $\nu=2J/\pi\hbar$.  Note that the initial state is chosen to
be the state with the lowest possible energy given the filling $n$ and
the population difference $n_R-n_L$. From the results for $I(\Delta
n)$ we obtain the current-voltage dependence $I(\Delta\mu)$ via
Eq.~(\ref{eqn:mu}).

\section{Atomtronic Diodes}
\label{sec:diode}

A diode is a circuit element which features a highly asymmetric
current-voltage curve. It allows a large current to pass in one
direction, but not in the other.  Thus, the analog of a diode is an
atomtronic circuit element that lets an atomic current pass through
when applying a voltage $V=\mu_L-\mu_R$ while $V=-(\mu_L-\mu_R)$ does
not generate a current or only a small saturation current.

\begin{figure}

~\hskip -0.5cm\includegraphics[width=8.5cm]{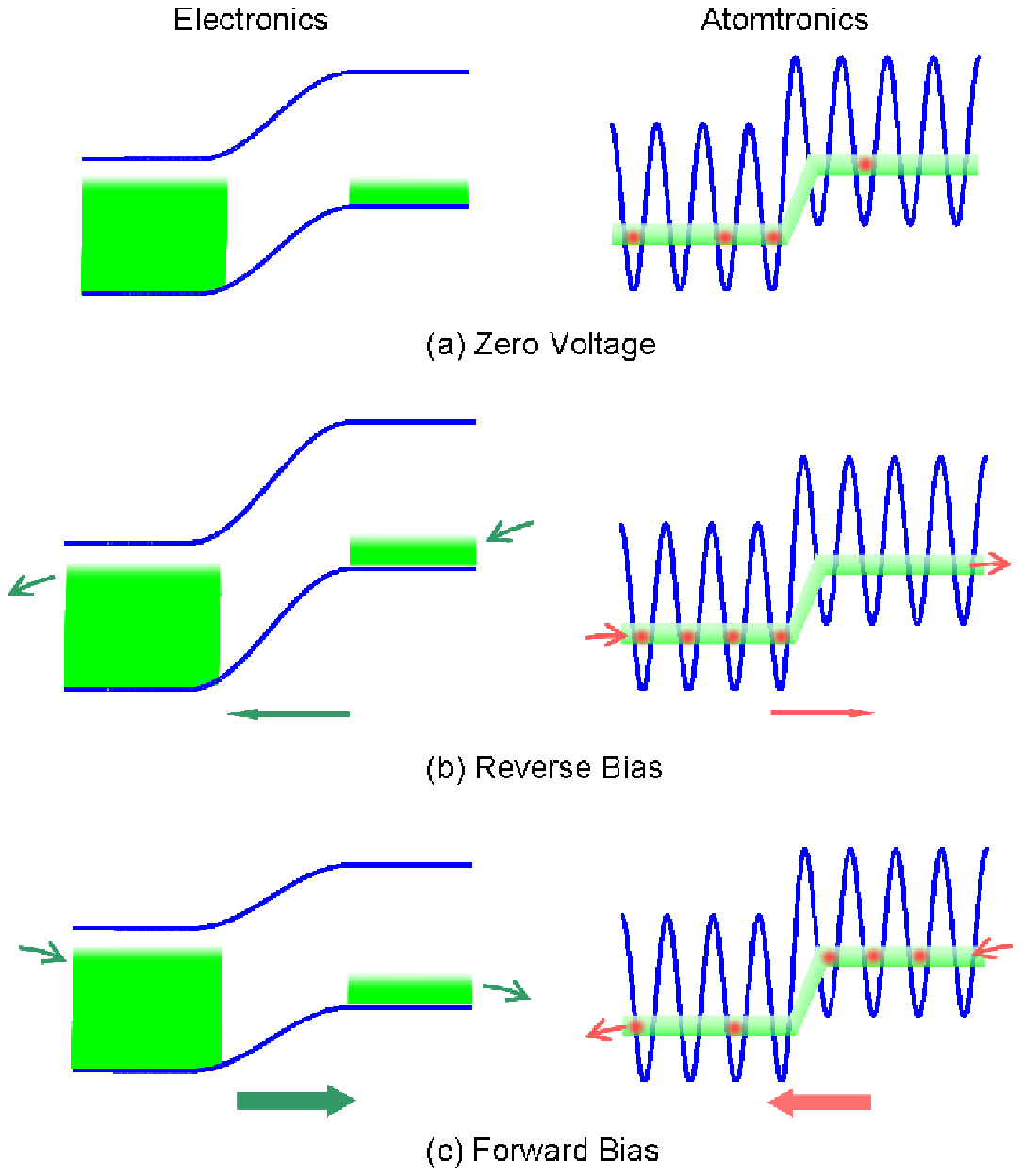}
\caption{Schematic of the conduction band of an electronic (left
panels) and atomtronic (right panels) PN-junction diode in (a)
equilibrium, (b) reverse bias and (c) forward bias with N-type
materials on the left and P-type on the right of each configuration.
\\ Left panels: The electronic system features a voltage dependent
energy barrier at the junction leading to an increasing (decreasing)
flux from the N-type to the P-type material as the junction is forward
(reverse) biased while the current from P to N is independent of
voltage.  \\ Right panels: The operation of an atomtronic diode is
based on the existence of insulating phases where $\partial n/\partial
\mu=0$, i.e. a change in voltage does not lead to a change in particle
transfer between battery and system. As a consequence, only a small
current can flow from N-component to P-component through an atomtronic
diode in reverse bias while in forward bias the particle transfer
between battery and system can be varied over a large range.
\label{fig:diodeschematic}
}
\end{figure}

In solid state electronics, diodes are built by setting up a
PN-junction in which a P-type semiconductor is brought into contact
with an N-type semiconductor. Electrons move through the junction
until an equilibrium is reached.  This process depletes the junction
region of free charges, leaving behind the static charges of the donor
and acceptor impurities. As illustrated in
Fig.~\ref{fig:diodeschematic}(a), this creates an effective potential
step across the junction.  When a reverse bias voltage is applied the
energy barrier is increased reducing the flux of electrons from N to
P. At the same time the number of electrons that can fall down the
step remains constant giving rise to a reverse bias saturation current
that is independent of voltage (see Fig.~\ref{fig:diodeschematic}
(b)).  However, if the diode is forward biased, more electrons are
able to move from the N-type to the P-type material than at
equilibrium since the potential barrier is decreased by the forward
bias voltage (see Fig.~\ref{fig:diodeschematic}(c)). A detailed
discussion of the diode behavior of a semiconductor PN-junction can
for example be found in \cite{Ankrum1971}.

\subsection{Diode PN-junction configuration}

As discussed previously, we can design atomtronic wires whose
conductivity properties can be described by locating the material's
chemical potential in its phase diagram. The properties of the
material can be changed by either changing its chemical potential or
by modifying the lattice and thereby altering the phase diagram. As
particular examples, we have discussed the analog of P- and N-type
doped conductors.

Materials which are composed of several types of atomtronic conductors
can be produced by connecting lattices of different doping. 
Another possibility to build a junction is to superimpose additional
external potentials, for example a simple potential step. This has the
effect of shifting the phase diagram of a part of the lattice upwards
or downwards with respect to that of the rest.  Sufficiently far away
from the junction, the state of the different components can be
accurately described by the phase diagrams of the individual
materials. In fact, we have found that the boundary only affects one
to two sites of each component in the limit $J/U\ll 1$.  This means
that a local chemical potential can be associated with each component
of the conductor which can be located in each material's phase
diagram.  Thereby, the local conductivity properties can be
identified. Of course, at equilibrium, i.e. at zero voltage, the
composite material is actually described by a single chemical
potential $\mu$ but it is $\mu$ relative to the zero point energy of
each lattice site that determines the filling of each site.

\begin{figure} \begin{center}
~\vskip -0.5cm\includegraphics[width=7.8cm]{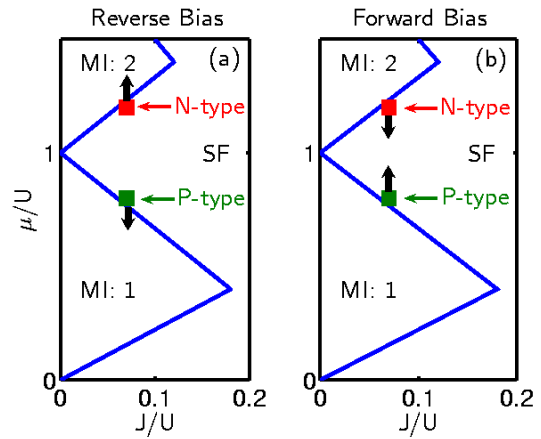}
\caption{Phase diagram of the PN-junction configuration of an
atomtronic diode. The small population of the second band in the
P-material yields the analog of the small thermal electron population
of the conduction band in a semiconductor. The states of the P-type
and N-type materials at zero voltage are represented by the
squares. Arrows indicate the chemical potential difference (voltage)
imposed to obtain a reverse (left) and a forward biased junction
(right).
\label{fig:pnjunctionvoltage}}
\end{center} \end{figure}

The conduction band of a semiconductor PN-junction features a
small thermal electron population on the P-side and a considerably
larger electron filling on the N-side. The atomtronic analog can be
obtained by bringing into contact P-type and N-type lattices.
Alternatively, it can be attained by imposing an external potential
step. In the following, we focus on this latter implementation. The
main characteristics of the diode behavior are not affected by this
choice.  An example for a possible equilibrium configuration is
represented by the squares in Fig.~\ref{fig:pnjunctionvoltage}. The
potential step could be generated experimentally by exposing one part
of the system to off-resonant laser light.

\subsection{Diode current-voltage characteristics}

To achieve diode characteristics, we exploit the possibility of
ndergoing a quantum phase transition to the insulating phase.  The
materials are configured such that the chemical potentials of the
battery poles remain in the superfluid regime when hooking up the
battery in one direction, but they easily enter insulating regimes
when the voltage is applied in the opposite direction.

The effect of applying a voltage is illustrated in
Fig.~\ref{fig:pnjunctionvoltage}. Forward bias is achieved by
connecting the P-side to the low voltage pole and the N-side to the
high voltage pole of the battery.  In this situation, the chemical
potentials of the battery poles are located in the superfluid region
of the phase diagram and atoms can flow from the P-component to the
N-component. The larger the applied voltage, the larger the generated
current. When the battery contacts are switched, the diode is reverse
biased. As the voltage is increased a small current starts
flowing. However, as soon as the voltage is large enough to make the
battery chemical potentials enter the insulating zones, the current
can not increase any further.  As a consequence, the current-voltage
curve is asymmetric. The remnant current obtained in reverse bias, the
saturation current, becomes smaller as the components' initial states
are moved closer to the insulating phase.

Figure \ref{fig:diodeschematic} presents a schematic comparison of the
conduction band of an electronic and an atomtronic diode obtained
using a step potential. For the atomtronic diode, the fact that only a
small current can flow in reverse bias is not due to the presence of a
voltage-dependent energy barrier at the junction as in the electronic
case. Instead, it arises from the battery chemical potentials moving
into insulating zones corresponding to a full conduction band on the
N-side and an empty conduction band on the P-side. An important
difference between electronic and atomtronic case is the opposite
direction of current flow. In forward bias, atoms flow from the P-type
to the N-type material rather than the other way around as electrons
in a semiconductor.

\begin{figure}
\begin{center}
\includegraphics[width=7.8cm]{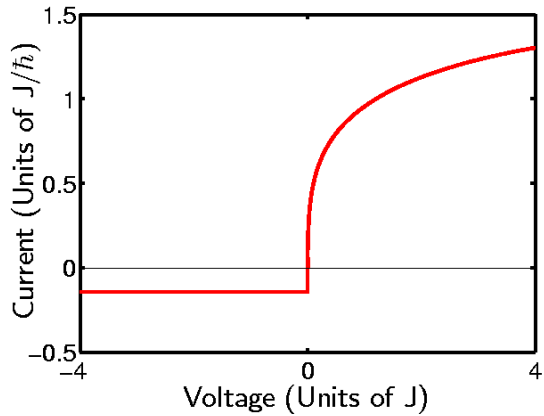}
\caption{The characteristic current-voltage curve for an atomtronic
diode. The larger the forward bias (voltage $>0$), the higher the
current of atoms flowing from the P-component and to the
N-component. In reverse bias (voltage $<0$) the current saturates
since the particle transfer between battery and system can not be
increased beyond a certain small value when the battery pole chemical
potentials enter insulating zones where $\partial n/\partial \mu=0$.
\label{fig:diode}}
\end{center}
\end{figure}

Figure \ref{fig:diode} displays the highly asymmetric current-voltage
curve obtained from our calculation. The potential step is chosen such
as to yield an equilibrium configuration with a filling of $1.99$ and
$1.01$ atoms on the N-side and P-side respectively. In this
configuration, we obtain a saturation current of $0.14\;J/\hbar$ while
in forward bias, currents can exceed $1.4\;J/\hbar$.  Note that the
current changes strongly in the vicinity of $V=0$. This is reminiscent
of the behavior of an electronic diode as the temperature approaches
zero.  Reducing the potential step lowers the population difference
between P- and N-component at equilibrium and leads to an increase in
saturation current and to a decrease of the slope of the
current-voltage curve around $V=0$.

The diode currents have been calculated analytically in the same
manner as the currents carried by atomtronic wires (see section
\ref{subsec:materialcurrentproperties} above) except for the addition
of the potential energy step between the two sites. From the result
for the relation $I(\Delta n)$ we obtain the current-voltage
dependence $I(\Delta \mu)$ using Eq.~(\ref{eqn:mu}).

\section{Atomtronic Transistor}
\label{sec:transistor}

\begin{figure}
\begin{center}
\includegraphics[width=7.8cm]{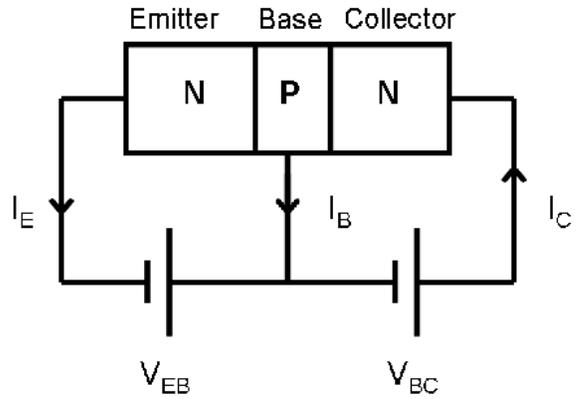}\\
\caption{Circuit schematic of an electronic or atomtronic bipolar
junction transistor of the NPN-type. A thin P-type component (base) is
sandwiched between two N-type components (emitter and collector).  The
key feature is that the voltage $V_{EB}$ can be used to obtain gain in
the collector current $I_C$ relative to the base current $I_B$. }
\label{fig:transistorcircuit}
\end{center}
\end{figure}

As in electronics, the highly asymmetric current-voltage curve of
atomtronic diodes can be exploited to build a transistor. Bipolar
junction transistors (BJT) are circuit elements can serve as
amplifiers or switches. In electronics, they consist either of a thin
P-type layer sandwiched between two N-type components (NPN) or a thin
N-type layer between two P-type components (PNP). For our discussion,
we consider an NPN transistor. A detailed discussion of semiconductor
bipolar junction transistors can, for example, be found in
\cite{Ankrum1971}.  The basic circuit schematic is displayed in
Fig.~\ref{fig:transistorcircuit}. The voltage $V_{BC}$ which is
applied to the PN-junction formed by the middle component, the base,
and one of the outer components, the collector, puts this junction
into reverse bias. At the same time, the other junction formed by the
base and the other outer component, the emitter, is put into forward
bias by applying a voltage $V_{EB}$. The key idea is to use the
voltage $V_{EB}$ to control the current $I_C$ leaving the collector
element and thereby achieve gain in $I_C$ relative to the base current
$I_B$.  At $V_{EB}=0$ one is simply dealing with the reverse biased
base-collector junction. In this case, the currents $I_C$ and $I_B$
both equal the small reverse bias saturation current of the
base-collector junction. The collector current $I_C$ grows drastically
when $V_{EB}$ is increased such that the emitter-base junction is
forward biased. This effect relies on the base region being very
thin. The forward bias gives rise to a flow of electrons from the
emitter into the base region, thereby significantly increasing the
number of electrons at the base-collector junction. Recall that at
$V_{EB}=0$, the base is depleted of electrons by the reverse bias
$V_{BC}$. The emitter-base junction thus serves to greatly modify the
number of carriers in the base which are subjected to the
base-collector reverse bias.  This leads to an increase of $I_C$
beyond the saturation current. Since the base is extremely thin there
is less opportunity for the electrons to leave the base compared to
entering the collector.  Because of this, most of the current that
enters the base from the emitter moves on to the collector instead of
it leaving out the base terminal.  Therefore the relative changes in
the current from the base $I_B$ and from the collector $I_C$ yields a
large differential gain $dI_C/dI_B$.

\begin{figure}
\begin{center}
\includegraphics[width=8cm]{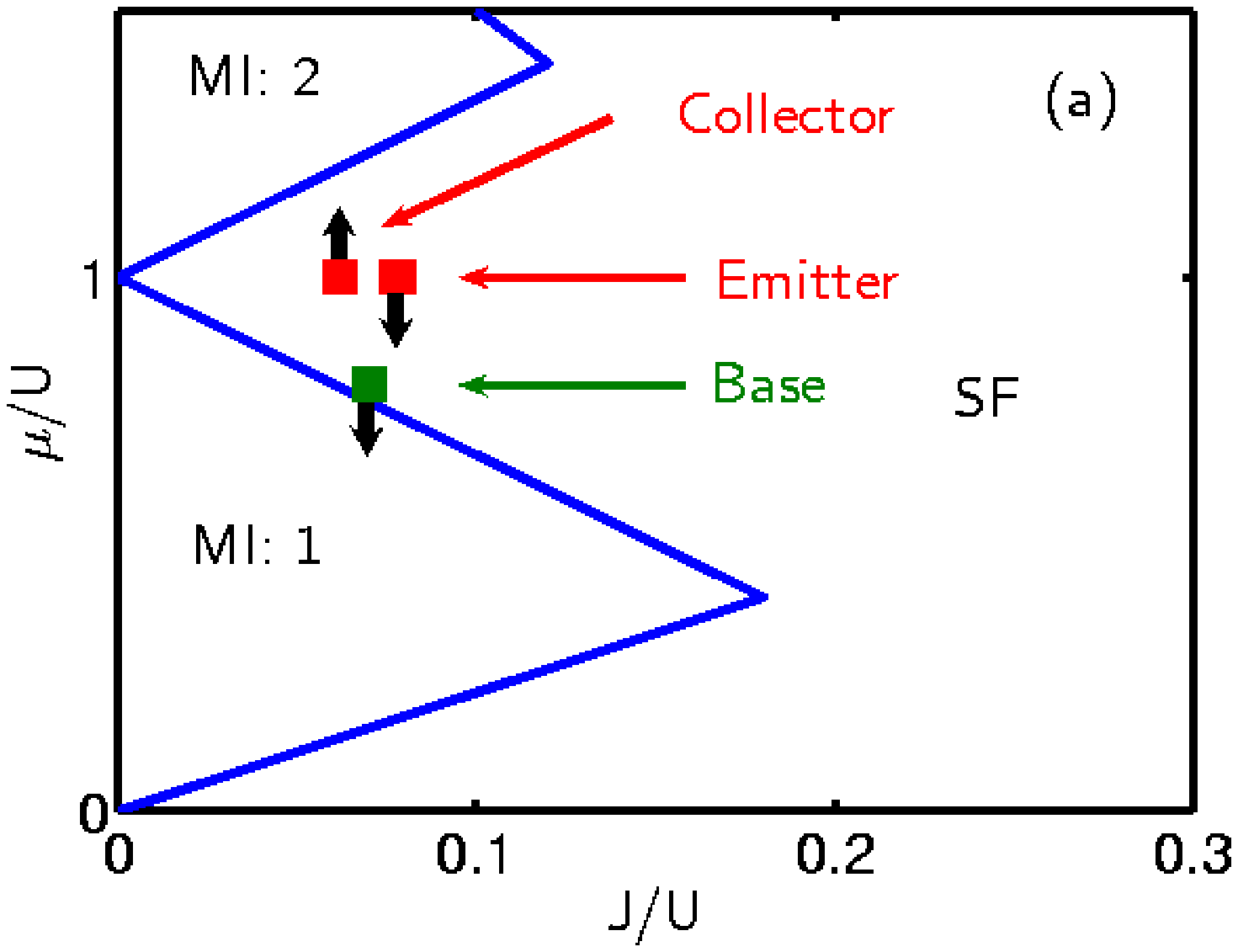}\\
~\hskip -0.67cm\includegraphics[width=7.8cm]{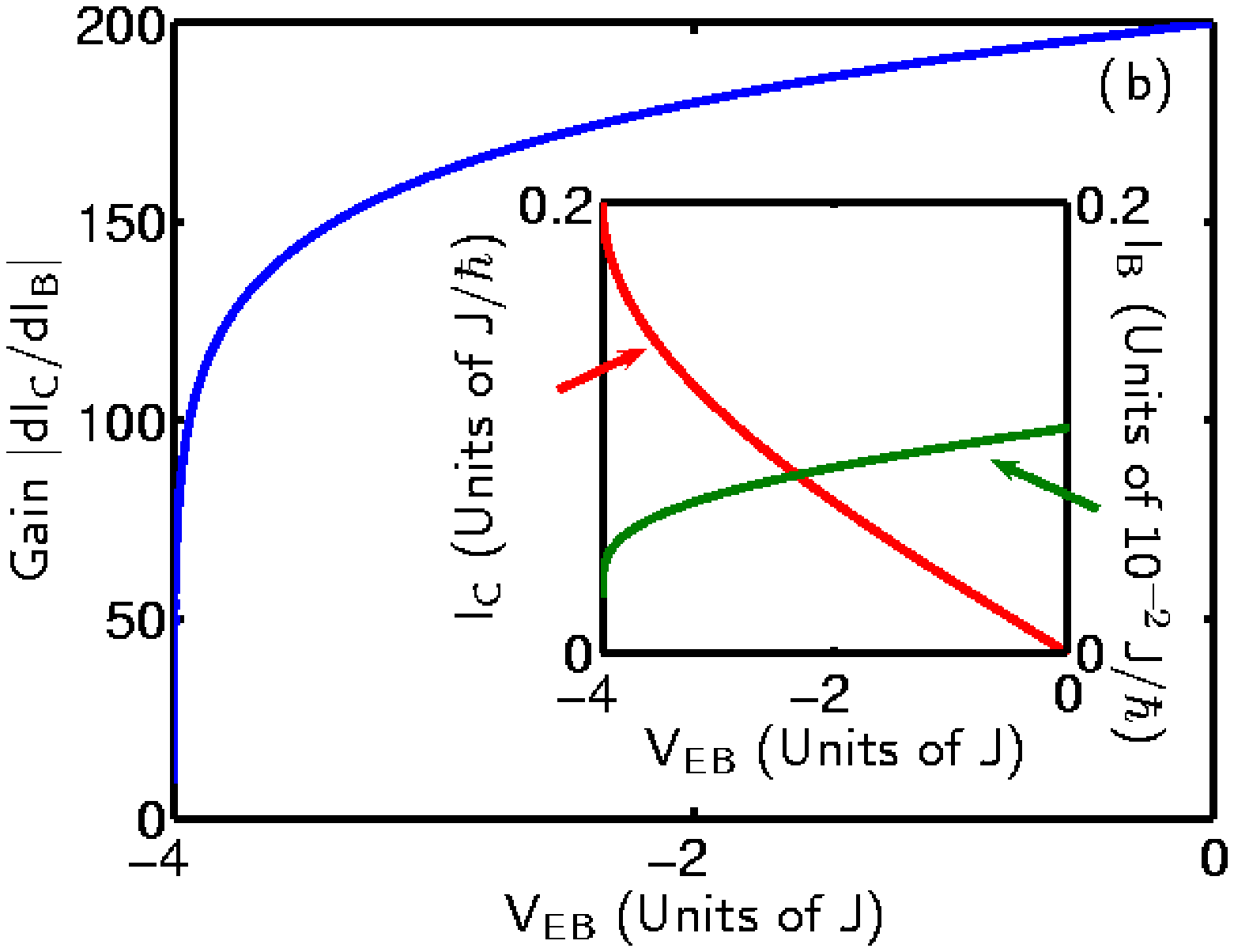}
\caption{(a) Squares: The equilibrium configuration of an atomtronic
bipolar junction transistor in the phase diagram. Arrows: Direction in
which the chemical potentials of the battery contacts are varied.  \\
(b) Differential gain $dI_C/dI_B$ as a function of emitter-base
voltage $V_{EB}$.  \\ Inset: Base current and collector current as a
function of emitter base voltage. The voltage is changed by varying
the emitter contact potential from $\mu(n_E=1)$ to $\mu(n_E=1.5)$. The
base contact is kept at $\mu(n_B=1)$ right below its equilibrium value
$\mu(n_B=1.005)$ while the collector contact is kept at the
equilibrium value $\mu(n_C=1.5)$.
\label{fig:transistor}}
\end{center}
\end{figure}

These key features of an electronic transistor can be translated into
atomtronics with atoms taking the place of electrons. The
important point is to mimic the carrier densities and thus the filling xs
of the three components, emitter, base and collector. This can be
achieved by doping or, equivalently, by using potential steps. As for
the diode case, we focus on the latter implementation.  The atomtronic
transistor can be created by setting up a configuration such that for
zero voltage the left and right regions have a large filling compared
to the sandwiched thin base region. The phase diagram for such an
arrangement is depicted in Fig.~\ref{fig:transistor}(a), where the
squares represent the equilibrium configuration, while the arrows indicate
the way the battery chemical potentials are tuned.  A small voltage
is applied to the collector-base junction such that a small current
flows from the collector into the base.  When the emitter battery
contact chemical potential $\mu_E$ is lowered, atoms move from base to
emitter and leave through the emitter, giving rise to a non-zero
emitter current $I_E$.  As a back-effect, this leads to an increase in
$I_C$ since the fast removal of atoms from the base through the
emitter allows more atoms to move into the base from the collector.

Meanwhile, the base current $I_B$ becomes smaller the further the
emitter chemical potential $\mu_E$ is lowered.  This is due to the
base being very thin relative to the emitter and the collector, so
atoms traverse preferentially from collector to emitter rather than
leaving out the base contributing to $I_B$.  The effect of the forward
bias $V_{EB}$ on the base current $I_B$ is thus opposite to its effect
on $I_C$.  Therefore, our atomtronic transistor features an inverted
amplification in which a small decrease in the base current goes along
with a large increase in the collector current (negative gain).

An example for the behavior of the currents in an atomtronic
transistor is given in Fig.~\ref{fig:transistor}(b). The equilibrium
configuration has $1.5$ atoms per site in emitter and collector and
$1.005$ atoms per site in the base. In Fig.~\ref{fig:transistor}(b)
we plot data obtained for the individual currents $I_C$, and $I_B$
upon variation of the emitter chemical potential $\mu_E$ while keeping
the base battery contact at a chemical potential $\mu(n_B=1)$ in the
$n=1$ Mott-insulating zone.  To demonstrate differential gain we
display the quantity $|dI_C/dI_B|$ as a function of $V_{EB}$.

The data presented in Fig.~\ref{fig:transistor}(b) is obtained from
the dynamics of a three-well lattice. An external potential is added
which raises the middle site. This site plays the role of the base
while the outer sites represent emitter and collector. The initial
state is prepared as the lowest energy solution with $n_E$ atoms in
the emitter and $n_C=1.5$ atoms in the collector. The current of atoms
passing through the base from collector to emitter is given by
$i_C=(i_{CB}+i_{BE})/2$ with $i_{CB}$ the current from collector to
base and $i_{BE}$ the current from base to emitter.  As in the
previous sections, we calculate the steady state current $I_C$ by
maximizing the time average of $i_C$.  From the result for the
relation $I_C(\Delta n)$ we obtain the current-voltage dependence
$I(\Delta \mu)$ using Eq.~(\ref{eqn:mu}).

Since we are interested in $\mu_E$-dependence
of $I_C$ at fixed $V_{BC}$ the initial collector occupation is kept
fixed at $n_C=1.5$ while the emitter occupation $n_E$ is varied in the
range $1\le n_E<1.5$.  The base current is set to be $I_B=\Gamma
n_B$, where $\Gamma$ must satisfy the condition $\Gamma\ll
J/\hbar$. The quantity $\Gamma$ describes the weak coupling with the
battery at the base contact which is due to the thinness of the
base. The data for $I_B$ displayed in Fig.~\ref{fig:transistor}(b) is
obtained using $\Gamma=0.01 J/\hbar$. Note that our calculation
neglects the effect of $I_B$ on $I_C$. This is justified because the
weak coupling at the base contact implies that the reduction of $n_B$
by $I_B$ is small.  Note that, overall, the possibility for transistor
current gain relies on two factors. The base region must be thin and
the emitter must have an equilibrium filling significantly larger than
the base.

\section{Remarks}
\label{sec:remarks}

We have showed how strongly interacting ultracold bosonic gases in
periodic potentials can be used as conductors in a circuit and how
they can be used to build atom analogs of diodes and bipolar junction
transistors. From here, the implementation of an atom amplifier is
immediate. An atom amplifier is a device which allows control of a big
atomic current with a small one. The transistor presented above
directly serves this purpose since small changes in the base current
bring along large changes in the collector current. From here, it is
straight forward to conceive of more complex devices such as a flip
flop, a bistable device that uses cross negative feedback between two
transistors.

The similarity in qualitative behavior goes along with a number of
significant differences in the underlying physics.  Firstly, in the
atom case the band gap results from interactions rather than from
statistics as in electronics. Secondly, the atomic currents are
superfluid. As a consequence, the ratio between voltage and current
has the meaning of a dissipationless resistance. Further differences
arise in both diodes and transistors.  Our atomtronic diode does not
feature a depletion layer, i.e. it does not exhibit a
voltage-dependent energy barrier at the junction. The asymmetry in the
current-voltage curve results from voltage-sign dependent transitions
to an insulating phase. As a consequence, atoms flow from P to N in
forward bias rather than flowing, as in electronics, from N to P.
Note that this difference can not be resolved by drawing the analogy
between atom holes and electrons rather than atoms and electrons since
this would also require relabeling the N-type material as P-type and
vice versa and hence the current direction would again be reversed in
comparison to the electronic case.  Due to the difference in diode
behavior the atomic collector current in a transistor flows from
collector into base and the emitter current flows from base into
emitter, i.e. opposite to electronic flow in a NPN transistor.  A
significant difference in the qualitative behavior of electronic and
atomtronic transistor is given by the gain being negative in the
atomtronic case. The collector current increases as base current
decreases.  We expect that this does not affect the functionality of
devices based on the operation of bipolar junction transistors. Yet,
an adaptation of their design will be necessary.

The data presented in this paper is obtained from calculations for a
one-dimensional lattice. This choice is of an entirely practical
nature. The basic ideas also hold for two- and three-dimensional cubic
lattices and extend to other lattice geometries which make transitions
between superfluid and insulating phases upon changes of the chemical
potential.

An issue to be addressed in the future is that of the noise associated
with the inherent quantum uncertainty of the current carrying states
given the context of the coherent transport of atoms.

Finally, it is important to keep in mind that this paper develops
atomtronics within the Bose-Hubbard model. This model provides an
excellent description of ongoing experiments with ultracold bosonic
atoms in optical lattices. Other Hamiltonians might offer alternative
ways of drawing the analogy with electronics. A natural choice for
further study are Hamiltonians describing bosons with beyond onsite
interactions and Hamiltonians for fermionic gases.

\section{Acknowledgments}
\label{sec:acknowledgments}

We thank John Cooper, Rajiv Bhat and Brandon M. Peden for useful
discussions. This work was supported by the Defense Advanced Research
Projects Agency's Defense Science Office through a PINS program (Grant
No. W911NF-04-1-0043) and the Air Force Office of Scientific Research
(Grant No. FA9550-04-1-0460). We also acknowledge support of the
Department of Energy, Office of Basic Energy Sciences via the Chemical
Sciences, Geosciences, and Biosciences Division, furthermore of the
National Science Foundation (B.~T.~S.) and the Deutsche
Forschungsgemeinschaft (M.~K.).



\begin{thebibliography}{10}
\bibitem{note}{The lower/upper boundary of the $m$th Mott zone is
obtained by applying the two-state approximation to the
$m$th/$(m+1)$th band, where $m=1,2,\dots$ is the band index. The tip
of the zone, given by the intersection of the two boundaries, yields a
good estimate of the value of $J/U$ at which the superfluid-insulator
transition occurs at integer filling.}
\end{thebibliography}
\end{document}